\RequirePackage{fix-cm}
\documentclass[smallextended]{svjour3}       % onecolumn (second format)
\smartqed  % flush right qed marks, e.g. at end of proof
\usepackage{graphicx}
\begin{document}

\title{Comment on Energy-Time Uncertainty Relations in Quantum Measurements}
\author{Koji Yasuda}

\institute{Institute of Materials and Systems for Sustainability, Nagoya University,\at
 Furo-cho, Chikusa-ku, Nagoya, 464-8601, Japan \\
	      \email{yasudak@imass.nagoya-u.ac.jp}
}

\date{Received: date / Accepted: date}

\maketitle

\begin{abstract}
A measurement apparatus analyzed in [Found. Phys. 46, 1522-1550 (2016)] does not exist.
\keywords{Quantum measurements \and Energy-time uncertainty relations \and Quantum information}
\end{abstract}

\section*{Introduction}
\label{intro}

In the paper \cite{miyadera2016}, properties of a quantum system were examined that acted as a measurement apparatus A, as well as the switching device of the interaction between A and another quantum system S to be measured. The apparatus is expected to simplify the problem of quantum measurements. A density matrix of the total system evolves under the time-independent Hamiltonian, $H = H_0 + V$, $H_0 = H_S + H_A$, where $H_A$ ($H_S$) is the Hamiltonian of the apparatus (system), and $V$ is the interaction between the system and apparatus. By selecting the Hamiltonian and the density matrix of the apparatus, $H_A$ and $\sigma$, and the system-apparatus interaction $V$, we want to satisfy the condition, $f(t)=0$ for $t \le 0$ and $f(t) \ne 0$ for $t > 0$, where
\begin{equation}
f(t) = \exp(-\frac{i}{\hbar} H t) (\rho \otimes \sigma) \exp(\frac{i}{\hbar} H t) - g(t), \label{eq1}
\end{equation}
\begin{equation}
g(t) = \exp(-\frac{i}{\hbar} H_0 t) (\rho \otimes \sigma) \exp(\frac{i}{\hbar} H_0 t).
\end{equation}
This is equivalent to Conditions 1 and 2 and Lemma 1 in the paper (without loss of generality we set $t_0 = 0$). The functions $f(t)$ and $g(t)$ represent the infinite-dimensional matrices, and $\rho$ is an arbitrary density matrix of the system. 
In this note we show that it is very hard to satisfy the condition and thus, such apparatus does not seem to exist.

Firstly, let us assume that the matrices $f(\tau)$ and $g(\tau)$ are well-defined  for complex $\tau = t + i \hbar \beta$ in the region $|\tau| < r$.
\begin{equation}
f(\tau) = e^{\beta H} f(t) e^{-\beta H} - g(\tau)
\end{equation}
\begin{equation}
g(\tau) = e^{\beta H_0} g(t) e^{-\beta H_0}
\end{equation}
This would be a resonable assumption, because $f(t)$, $g(t)$ and the statistical operators $e^{- \beta H}$, and $e^{- \beta H_0}$, are well-defined by physical requirement. 
The continuity of $g(\tau)$ is verified easily as
\begin{equation}
\lim_{\delta \rightarrow 0} g(\tau + \delta ) = 
\lim_{\delta \rightarrow 0} e^{- i H_0 \delta / \hbar} g(\tau) e^{i H_0 \delta / \hbar} \rightarrow g(\tau).
\end{equation}
Similarly, $f(\tau)$ should be continuous.
Secondly, the first derivative,
\begin{equation}
-i\hbar \frac{df}{d\tau}(\tau) = [H, f(\tau)] + [V, g(\tau)],
\end{equation}
is well-defined and continuous in the same region. According to Goursat's lemma, higher order derivatives exist.
For example, the second derivative of the matrix elements are
\begin{equation}
(-i\hbar)^2 \frac{d^2f}{d\tau^2}(\tau) =[H, [H, f(\tau)] + [V, g(\tau)]] + [V, [H_0, g(\tau)]].
\end{equation}
Thus, we have the Taylor expansion of $f(\tau)$ about $\tau = 0$.
\begin{equation}
f(\tau) = f(0) + \frac{d f}{d \tau } (0) \tau + \frac{1}{2} \frac{d^2 f}{d \tau^2 } (0) \tau^2 + \cdots
\end{equation}
Since $f(t)=0$ for real and negative $t$ by assumption, all the derivatives in the expansion are zero. Thus we conclude that $f(t)=0$ for $0 < t < r$. 

% It was proved in the original paper that the total Hamiltonian must be two-side unbounded, namely, it neither has a minimum nor a maximum eigenvalue. This might be a different and indirect statement of the present result.

The analytic property of Eq.\ (\ref{eq1}) is apparent in matrix form.
Assuming that the system is in a large box, we have eigenvalues and normalized eigenfunctions of $H$ and $H_0$, denoted as ($E_I$, $\Psi^I$) and ($\epsilon_I$, $\Phi^I$), respectively. All the matrix elements of the equation,
\begin{equation}
f_{k l}(t) = \sum_{I J} \Psi_k^I e^{-i E_I t /\hbar} \left< \Psi^I \right| \rho \otimes \sigma \left| \Psi^J \right> e^{i E_J t /\hbar} \Psi_l^J - g_{k l}(t),
\end{equation}
\begin{equation}
g_{k l}(t) = \sum_{I J} \Phi_k^I e^{-i \epsilon_I t /\hbar} \left< \Phi^I \right| \rho \otimes \sigma \left| \Phi^J \right> e^{i \epsilon_J t /\hbar} \Phi_l^J ,
\end{equation}
are regular analytic functions of complex variable $t$ except for $t = \infty$. Thus every matrix element has the Taylor series about $t=0$, and the convergence radius is infinity. 

The author acknowledges Prof. Miyadera for his valuable discussions.

\end{document}